\documentclass[12pt,leqno,twoside]{article}
\usepackage{amsmath,amsfonts,amssymb,amsthm}
\usepackage{graphicx}
\usepackage{hyperref}

\newcommand{\be}{\begin{equation}}
\newcommand{\en}{\end{equation}}

\newcommand*{\pd}
[2]{\mathchoice{\frac{\partial#1}{\partial#2}}
  {\partial#1/\partial#2}{\partial#1/\partial#2}
  {\partial#1/\partial#2}}

\renewcommand{\vec}[1]{\boldsymbol{#1}}

\newtheorem{remark}{Remark}

\title{On the Mathematical and Geometrical Structure \\ of the Determining
  Equations for Shear Waves\\
  in Nonlinear Isotropic Incompressible Elastodynamics}
\author{Giuseppe~Saccomandi
  \\
\footnotesize\texttt{giuseppe.saccomandi@unipg.it}
\\\footnotesize Dipartimento di Ingegneria,
  \\
\footnotesize  Universit\`{a} degli Studi di Perugia, 06125 Perugia, Italy
\\[0.5cm]
Raffaele Vitolo
  \\
\footnotesize\texttt{raffaele.vitolo@unisalento.it}
\\
\footnotesize\texttt{http://poincare.unisalento.it/vitolo}
\\
\footnotesize Dipartimento di Matematica e Fisica ``E. De Giorgi'', \\
\footnotesize  Universit\`{a} del Salento, 73100 Lecce, Italy
}

\date{\today}

\begin{document}
\numberwithin{equation}{section}
\maketitle


\begin{abstract}
  Using the theory of $1+1$ hyperbolic systems we put in perspective the
  mathematical and geometrical structure of the celebrated circularly polarized
  waves solutions for isotropic hyperelastic materials determined by Carroll in
  Acta Mechanica 3 (1967) 167--181. We show that a natural generalization of
  this class of solutions yields an infinite family of \emph{linear} solutions
  for the equations of isotropic elastodynamics. Moreover, we determine a huge
  class of hyperbolic partial differential equations having the same property
  of the shear wave system. Restricting the attention to the usual first order
  asymptotic approximation of the equations determining transverse waves we
  provide the complete integration of this system using generalized symmetries.

\textbf{MSC 2010}: Elasticity, hyperbolic PDEs, symmetries

\end{abstract}

\maketitle
\newpage
\section{Introduction and Basic Equations}

The propagation of transverse or shear waves in
incompressible isotropic nonlinear hyperlasticity is governed by a
system of non-linear equations in $1+1$ independent variables $(x,t)$
\begin{align} \label{1} \notag
  & \varrho u_{tt}-\left[Q(u_x^2+v_x^2)u_x \right]_x=0,\\
  \\ \notag & \varrho v_{tt}-\left[Q(u_x^2+v_x^2)v_x \right]_x=0,
\end{align}
where $u=u(x,t)$ and $v=v(x,t)$ are the unknown functions (the transverse
motions), $\varrho$ is the constant density and $Q=Q(u_x^2+v_x^2)$ is the
generalized shear modulus\cite{DS2}.

It is usual to derive \eqref{1} with respect to $x$, and to
introduce as new unknowns the strains $U=u_x$ and $V=v_x$,
to rewrite \eqref{1} 
as a first-order homogeneous quasilinear
system
\begin{equation} \label{1hyp} \vec{U}_t + \mathcal{A}(\vec{U}) \vec{U}_x=0,
\end{equation}
see, for example, \cite{JT} or system ($7.1.14$) page $176$ in
\cite{Dafermos}.  Indeed, introducing $\tilde{Q}=Q/\varrho$ in terms of
the strains
\begin{equation}
\label{1bishyp}
U_t=M_x, \quad M_t=[\tilde{Q}U]_x, \quad V_t=N_x, \quad N_t=[\tilde{Q}V]_x.
\end{equation}

On the other hand, the highly symmetric structure of \eqref{1} suggest an
alternative compact form for this system.  Using the complex function $W$ with
modulus $\Omega$ and argument $\theta$, defined by
\begin{equation}
  \label{1tris}
  W(x,t)=\Omega(x,t) \exp[i\theta(x,t)]=U(x,t)+i V(x,t),
\end{equation}
so that
$$
U=\operatorname{Re}\{W\}=\Omega \cos(\theta), \quad
V=\operatorname{Im}\{W\}=\Omega \sin(\theta),
$$
we rewrite \eqref{1} as a single complex equation
\begin{equation}
  \label{2}
  \varrho W_{tt}-\left[Q(\Omega)W\right]_{xx}=0.
\end{equation}

This complex format has been used in the study of the nonlinear string by
Rubin, Rosenau and co-workers\cite{Ros3, Ros1, Ros2} and by Destrade and
Saccomandi in nonlinear elasticity and some related theories\cite{DS,DS2,
  DS1}, to obtain some similarity reductions for the system \eqref{1} and
several exact solutions. This format has been fundamental to unify all of the
Carroll's results\cite{C, C1, C2} and to generalize them.  Moreover, this
format has been used by Rogers\cite{Rogers} to show a relationship between
\eqref{1} and the Ermakov--Ray--Reid system.

For example, let us consider a special class of such similarity solutions: the
Carroll's finite amplitude circularly-polarized harmonic progressive
waves\cite{C}.  These solutions are obtained considering $\Omega=A$ and
$\theta=kx-\omega t$ and are defined as
\begin{equation} \label{13bis}
  U(x,t)=A \cos(kx-\omega t), \quad V(x,t)=\pm A
  \sin(kx-\omega t),
\end{equation}
where the amplitude $A$, the wave number $k$, and the frequency $\omega$ are
real positive constants, and the plus (minus) sign for $V$ corresponds to a
left (right) circularly-polarized wave.  The waves \eqref{13bis} are solutions
for the system \eqref{1} if the dispersion relation
\begin{equation}\label{13}
  \varrho \omega^2=k^2Q(A^2).
\end{equation}
is satisfied.  Under very mild conditions on the generalized shear modulus
\eqref{13} is clearly satisfied and therefore \eqref{13bis} are an example of
smooth global solutions of the Cauchy problem in the whole space composed by 
\eqref{1} and the initial conditions
\begin{align*}
  & u(x,0)=A \cos(kx), &  &v(x,0)=\pm A \sin(kx),
  \\
  & u_t(x,0)=A \omega \sin(kx), &  &v_t(x,0)=\mp A \omega \cos(kx).
\end{align*}
The \eqref{13bis} are interesting not only because they are beautiful exact
closed form solutions for a large class of materials but, to the well educated
reader, it is clear that these solutions are \emph{exceptional} in their
mathematical character: global solutions of permanent form of a nonlinear
hyperbolic system.

The possibility of smooth global solution for equations \eqref{1} has been
first noticed by John in his celebrated 1974 paper\cite{Fritz}:
\begin{quote}
  \emph{On the other hand, there are also known special non-singular wave
    solutions (e.g., the transverse waves in Hadamard materials (see $[4]$), or
    the Caroll waves in arbitrary materials (see $[6]$).  This raises the
    question whether the results of the present paper that depend on
    \emph{genuine non-linearity} of the system ever apply to plane elastic
    waves. It will be proved here that there is indeed a large class of plane
    waves where we have no genuine non-linearity, namely all those in which the
    wave front contains a principal direction of strain.}
\end{quote}
John is pointing out that there is a special class of materials (Mooney-Rivlin
materials in the incompressible setting) for which \eqref{1} reduces to a
linear system of equations. (For Mooney-Rivlin materials we have $Q \equiv \mu$
where $\mu$ is the constant infinitesimal shear modulus).

On the other hand, for general class of materials,
the special structure of the matrix $\mathcal{A}$ in \eqref{1hyp} reduces the
problem to one involving the ordinary simple waves of a homogeneous system
comprising only two equations. Each simple wave is associated with one of the
four eigenvalues\footnote{The plus and minus sign are associated to forwards
  and backwards waves respectively.} $\lambda_{\pm}^{(1)}$ and
$\lambda_{\pm}^{(2)}$ of $\mathcal{A}$.  Being $\lambda^{(1)}>\lambda^{(2)}$,
using a terminology introduced in \cite{Collins}, it is usual to refer to
such waves as \emph{fast} and \emph{slow} waves.

Slow waves are \emph{exceptional} or \emph{linearly degenerate} in the
hyperbolic systems language, i.e. the gradient of the corresponding eigenvalue
is orthogonal to the corresponding eigenvector.  It is well known (see
corollary $8.2.6$ page $211$ in \cite{Dafermos}) that when a
characteristic family of an hyperbolic system is linearly degenerate travelling
waves solutions are possible. This corollary contains the \emph{mathematical
  reason} of the existence of Carroll waves.

We record several papers concerning linearly degenerate or exceptional systems
in elastodynamics. Some papers are dedicated to existence theorems, for example
\cite{Liu}, others, mainly by the Boillat's school \cite{Boi1, Boi2,
  Donato1}, are more focused on the interplay between this peculiar
mathematical structure and the constitutive nature of some material laws. The
aim of such papers is to determine special classes of elastic strain-energy
densities for which the mathematical resolution is simpler than usual because
the equations are completely exceptional.  Moreover, there is a huge literature
about plane transverse acceleration waves in elastic solids (see for example
\cite{Scott}), where it is noticed that in incompressible materials there
is a direction where waves may propagate without a change in amplitude because
of the exceptional character of one of their eigenvalues.

On the other hand, these exact solutions must play a role in the framework of
the well-posedness of the Cauchy problem for incompressible dynamic
elasticity. This subject is denoted as problem number $12$ in John Ball's
review about open problems in nonlinear elasticity\cite{Ball}\footnote{The
  problem $12$ in \cite{Ball} is: \emph{Prove the global existence and
    uniqueness of solutions to initial-boundary value problems for properly
    formulated dynamic theories of nonlinear elasticity.}}.  In reviewing this
problem Ball cites the works by Ebin and coworkers\cite{E1, E2, E3} and the
work by Hrusa and Renardy\cite{Renard}.  The paper \cite{E2} (dated 1996)
opens in a very significant, communicative and direct way:
\begin{quote}
  \noindent \textbf{Theorem.} \emph{The initial value problem for the equations
    of motion of an incompressible hyperelastic homogeneous isotropic material
    has classical solutions for all time, if the initial displacement and
    velocity are small.}
\end{quote}

It is clear that in \cite{E1, E2, E3, Renard} the investigations are not
restricted to transverse waves, but clearly Carroll's solutions
are an example of global
existence for large amplitude initial data in unbounded domains. 

The more recent paper by Sideris and Thomases\cite{Sideris} points out that
since in isotropic nonlinear elastic systems shear waves are linearly
degenerate the so called \emph{null condition} is automatically satisfied. This
fact is used to confirm the \emph{intuitive idea} that suitable weighted local
decay estimates for the perturbative equations can be expected via the
generalized energy method and therefore the existence of global-in-time
classical solutions to the Cauchy problem for incompressible elastic materials
is proved for \emph{small} initial displacements. This fact is confirmed in
\cite{Lei} where the possibility to obtain such kind of results is
connected to the exceptionality of the equations.  In \cite{Lei} we read:
\begin{quote}
  \emph{The equations of incompressible elastodynamics display a linear
    degeneracy in the isotropic case; i.e., the equation inherently satisfies a
    null condition. By taking the advantage of this structure, we prove that
    the 2-D incompressible isotropic nonlinear elastic system is almost
    globally well-posed for small initial data.}
\end{quote}
Once again the connection between the linearly degenerate structure of the
equations and the possibility to find global solutions for the Cauchy
problem (and this also for large data) seems to have been not noticed. 

This situation is strange because the paper \cite{Lei} focuses on the
neo-Hookean material in the two dimensional case. (We point out that in the two
dimensional case the neo-Hookean material cannot be distinguished from the
Mooney-Rivlin material.) The fact that many solutions of the neo-Hookean model
may be found solving linear equations is well known to the practitioners of
nonlinear elasticity, see for example \cite{Lei0}.

The aim of the present note is to push forward the connection between the
exceptionality of the system \eqref{1} and the existence of exact solutions as
\eqref{13bis}.  In so doing we determine explicitly a huge class of smooth
solutions.  Moreover, we are able to point out that this kind of exact
solutions are peculiar of an entire class of linearly degenerate second order
differential equations.

The plan of the paper is the following. In Section $2$ we generalize the
\eqref{13bis}.  In so doing, we provide a huge class of new exact solutions for
the equations of non-linear elastodynamics.  In Section $3$ with derive the
usual first order asymptotic model associated with \eqref{1}. This allows to
introduce a simpler format for our investigations. We realize that the
asymptotic system corresponding to \eqref{1} is a Temple system\cite{T,AF}.
Using symmetry transformations we are able to provide not only the
full class of \emph{linear} solutions for such system, but to derive also the
general integral.  The last Section is devoted to concluding remarks.

Symbolic computations were performed in CDIFF\cite{cdiff}, a freely available
REDUCE\cite{reduce} package for computations in the geometry of differential
equations.

\section{A Generalization of Carroll Solutions}

Let us consider $\Omega=\text{const.}$ in the system \eqref{1tris} . In this
case starting from \eqref{2} we obtain an overdetermined system in the unknown
$\theta=\theta(t,x)$. The general solution of this overdetermined system, when
$Q >0$, is simple and given by the solutions of the first order wave equations
$$
\theta_{t}\pm\sqrt{Q/\varrho}\theta_{x}=0.
$$
This means that the Carroll's solutions \eqref{13bis} are only one possible
choice among infinite possibilities.

In general it is possible to have
solutions in the form
\begin{equation}\label{14}
  U(x,t)=A \cos(\theta), \quad V(x,t)=\pm A \sin(\theta),
\end{equation}
where $\theta=F(x\pm \sqrt{Q/\varrho}t)$ and $F$ is an arbitrary function.  Of
course, new solutions cannot be obtained as sums of solutions $\theta$ with
different signs since they are solutions of nonlinear equations.  This is the
\emph{large class of plane waves where we have no genuine non-linearity}
identified in \cite{Fritz}. To the best of our knowledge, the explicit
determination of such exact solution have been unnoticed.

If in \eqref{1tris} we set $\theta=\text{const.}$, i.e. we
consider a plane polarized wave, from \eqref{2} we obtain the single real
second order partial differential equation $\varrho
\Omega_{tt}-[Q(\Omega)\Omega]_{xx}=0$ for which, if $Q\neq\text{const.}$, any
solution blows up\cite{Kla}.
 
We point out that to ensure that the generalized shear modulus $Q$ is
positive it is sufficient to impose that the strain-energy density satisfies the
usual \emph{empirical inequalities}\cite{Ant}.

The possibility to find explicitly a similar huge class of exact solution for a
nonlinear system of partial differential equations is not restricted to the
system \eqref{1}. Let us consider the abstract mathematical system
\begin{subequations}
  \label{1bis}
  \begin{align}
    & U_{tt}-\left[P(U,V)U \right]_{xx}=0, \\ & V_{tt}-\left[P(U,V)V
    \right]_{xx}=0,
  \end{align}
\end{subequations}
where $P$ is a suitable \emph{constitutive} function.  We suppose that
$P(U,V)>0$ in the domain of interest.  We can set $P(U,V)=A$ where $A>0$ is a
constant and therefore, under suitable assumptions on $P$, it is possible to
write $V=\Psi(U;A)$.  Then the system \eqref{1bis} is transformed in the
overdetermined system
\begin{subequations}
  \label{ppp}
  \begin{align}
    & U_{tt}-AU_{xx}=0,\\
    & [\Psi(U;A)]_{tt}-A\left[\Psi(U;A)\right]_{xx}=0.
  \end{align}
\end{subequations}
It is easy to check, by a direct computation, that $U=F(x\pm\sqrt{A}t)$ solves
this overdetermined system. On the other hand, the system \eqref{1bis} is
compatible if $V=kU$ where $k$ is an arbitrary constant and in this case we
reduce the system to a single genuinely non-linear partial differential
equation of the second order.

Therefore, we have pointed out a very general result peculiar to the system
\eqref{1bis} of second order hyperbolic differential equations in $1+1$
dimensions in two unknowns.  To the best of our knowledge the explicit
characterization of the solutions we have provided has never been noticed.

\begin{remark}
  If $P=P(U/V)$ the substitution $V=kU$ reduces the system
  \eqref{1bis} to a set of two uncoupled linear differential equation.  This
  choice of $P$ in the family \eqref{1bis} is special as we can check when
  $P=U/V$. This system is completely exceptional\cite{BT}.
\end{remark}

\section{An Asymptotic Model}
In nonlinear acoustics it is usual to derive an asymptotic model for the system
\eqref{1}. This Section is devoted to a detailed discussion of such a system.
The system has a mechanical interest, and its mathematical structure has been
deeply studied (see for example \cite{KK}).

Let us introduce the Taylor expansion of the generalized shear
modulus
$$
Q(U^2+V^2)=\mu_0+\mu_1(U^2+V^2) + \ldots ,
$$
and let us assume $U=\epsilon \hat{U}, \, V=\epsilon \hat{V}$ introducing the
new independent variables $X=\epsilon^2 x$ and $\tau=t-x/\mathfrak{c}_0$, where
$\mathfrak{c}_0=\mu_0/\varrho$. Here $\epsilon$ is a small parameter.
Considering only terms up to $\mathcal{O}(\epsilon^3)$ and introducing the
notation $U:=\partial \hat{U}/ \partial \tau, \, V:=\partial \hat{V}/ \partial
\tau $ we obtain the first order hyperbolic system\footnote{We point out that
  the system we derive is exactly the \emph{toy} hyperbolic system introduced in
  \cite{Dafermos} page $182$ formula ($7.2.11$).}
\begin{subequations} \label{51}
  \begin{align}
    & U_{X}-\beta \left[(U^2+V^2) U \right]_{\tau}=0,\\
    & V_{X}- \beta \left[(U^2+V^2) V \right]_{\tau}=0,
  \end{align}
\end{subequations}
where $\beta=\mathfrak{c}_1/(2\mathfrak{c_0^2})$ (here
$\mathfrak{c}_1=\mu_1/\varrho$). 

Introducing polar coordinates\footnote{here $\rho$ is not to be confused with
  the density $\varrho$.}
\begin{equation} \label{lineari} U=\rho(X,\tau) \cos \vartheta(X, \tau), \quad
  V=\rho(X,\tau) \sin \vartheta(X, \tau),
\end{equation}
the system \eqref{51} is rewritten as
\begin{subequations}\label{52bis}
  \begin{align}
    \label{52bisa} & \vartheta_X-\beta \rho^2 \vartheta_{\tau}=0,
    \\
    \label{52bisb} & \rho_X-3 \beta \rho^2 \rho_{\tau}=0 \rightarrow
    \rho_X-\beta (\rho^3)_{\tau}=0.
  \end{align}
\end{subequations}

If $\vartheta=\text{const.}$ we have plane polarized waves and the system
\eqref{52bis} reduces to a single partial differential equation whose exact
solution is $\rho=\Phi(\tau-3\beta X \rho^2)$.

On the other hand, a class of remarkable solutions for system \eqref{52bis} is
obtained when $\rho=A$, where $A$ as in the circularly polarized wave solutions
\eqref{13bis} is an arbitrary constant. This class of solutions is obtained
solving a linear equation, indeed, in this case we have the solutions
\begin{equation}
U=A \cos\left(\Theta(\xi)\right), \quad V=A \sin\left(\Theta(\xi)\right),
\label{eq:2}
\end{equation}
where $\vartheta=\Theta(\xi)$ and $\xi=\beta A^2X+\tau$.  This is an infinite
family of smooth solutions of the nonlinear system \eqref{51}.

The system \eqref{52bis} maybe easily rewritten in conservative form as
\begin{subequations} \label{52ter}
  \begin{align}
    \label{52tera} & (\rho \vartheta)_X-\beta (\rho^3 \vartheta)_{\tau}=0,
    \\
    \label{52terb} & \rho_X-\beta (\rho^3)_{\tau}=0.
  \end{align}
\end{subequations}
The general solution of the system \eqref{52ter} may be represented introducing
the potential variable $\phi=\phi(X, \tau)$ such that
\begin{equation}\label{potential}
  \phi_\tau=\rho, \quad \phi_X=\beta \rho^3.
\end{equation}
Indeed, the general integral (in implicit form) of \eqref{52terb} is well known
and for any given solution $\rho(X, \tau)$ it is possible to define a corresponding $\phi$
from \eqref{potential}. If we consider $\vartheta=F(\phi)$, where $F$ is an
arbitrary function, we obtain the general solution of \eqref{52bis}.  By a direct
check denoting $dF/d\phi=F'$ we compute
\begin{displaymath}
  (\rho F)_X-\beta (\rho^3 F)_{\tau} \equiv \rho F' \phi_X-\beta \rho^3 F'
  \phi_\tau \equiv  \rho F' \left( \phi_X-\beta \rho^2 \phi_\tau \right)
  \equiv 0.
\end{displaymath}

Another representation of the exact solution of this system (when $\rho \neq
\text{const.}$) is obtained by the so-called \emph{generalized hodograph
  method}\cite{ST}. The method makes use of families of commuting
generalized (or higher) symmetries (see, for example, \cite{Many,Olver})
of the system \eqref{52bis}. Such symmetries are generalized (or higher) vector
fields $\varphi^\vartheta \pd{}{\vartheta} + \varphi^\rho\pd{}{\rho}$. Here the
word ``generalized'' means that the coefficients are functions of derivatives
of an arbitrarily high order, \emph{i.e.}  $\varphi^\vartheta =
\varphi^\vartheta(X,\tau,\vartheta,\rho,\vartheta_X,\vartheta_\tau,
\rho_X,\rho_\tau,\dots)$ and analogously for $\varphi^\rho$. Generalized vector
fields are generalized symmetries if they are solutions of the linearized
system
\begin{subequations}
  \begin{align}
    & D_X(\varphi^\vartheta) - 2\beta\rho\vartheta_\tau \varphi^\rho -
    \beta\rho^2D_\tau(\varphi^\vartheta) = 0
    \\
    & D_X(\varphi^\rho) - 6\beta\rho\vartheta_\tau \varphi^\rho -
    3\beta\rho^2D_\tau(\varphi^\rho) = 0
  \end{align}
\end{subequations}
\emph{over} the system \eqref{52bis}. \emph{Hydrodynamic type symmetries} are
generalized symmetries that have the simplest structure with respect to
derivatives: for our equation they are of the form $\varphi^\vartheta =
\varphi^\vartheta(\vartheta,\rho)\vartheta_\tau$ and analogously for
$\varphi^\rho$. Note that a vector field on the space of dependent and
independent variables $\xi^\tau\pd{}{\tau} + \xi^X\pd{}{X} + \eta^\vartheta
\pd{}{\vartheta} + \eta^\rho\pd{}{\rho}$ is a classical (point) symmetry if and
only if its vertical part $(\eta^\vartheta - \vartheta_X\xi^X - \vartheta_\tau
\xi^\tau)\pd{}{\vartheta} + (\eta^\rho - \rho_X\xi^X
-\rho_\tau\xi^\tau)\pd{}{\rho}$ is a solution of the above linearized system; in
this sense the hydrodynamic type symmetries are the simplest generalized
symmetries.

It is known\cite{ST} that diagonal hydrodynamic-type systems
in $2$ dependent variables admit a space of hydrodynamic-type symmetries which
are parametrized by two arbitrary functions and commute, as vector fields, with
the vector field defined by the right-hand side of the differential equation.

In our case, the system~\eqref{52bis} admits the hydrodynamic symmetries
\begin{equation}
  \label{eq:35}
  \phi = - \left(\frac{s_3(\theta)}{\rho} + s_4(\rho)\right)\theta_\tau
  \frac{\partial}{\partial\theta}
  -\left(\frac{d(s_4(\rho)\rho)}{d\rho}\right) \rho_\tau
  \frac{\partial}{\partial\rho}
\end{equation}
The above symmetries commute with the vector field
$\varphi=$ $\beta\rho^2\vartheta_\tau \pd{}{\vartheta} +
3\beta\rho^2\rho_\tau\pd{}{\rho}$ which is given by the right-hand side of the
equation~\eqref{52bis} (see the Appendix for more
details).  Then every solution of the algebraic system
\begin{subequations}\label{eq:sol}
  \begin{align}
    \label{eq:sol1} &-\left(\frac{s_3}{\rho} + s_4\right) = \beta \rho^2 X +
    \tau
    \\
    \label{eq:sol2} &-\left(\frac{ds_4}{d\rho}\rho + s_4\right) = 3\beta \rho^2
    X + \tau
  \end{align}
\end{subequations}
(where $s_3=s_3(\theta)$ and $s_4=s_4(\rho)$ are two arbitrary functions) is a
solution of the system~\eqref{52bis} with the property that $u^i_\tau\neq 0$,
and conversely any solution of~\eqref{52bis} with the property $u^i_\tau\neq 0$
can be locally represented as a solution of~\eqref{eq:sol}. Indeed, the
system~\eqref{52bis} can be solved through the hodograph transformation
$(X,\tau)\mapsto (\theta,\rho)$. If one performs that transformation the
system~\eqref{52bis} becomes
\begin{subequations}\label{eq:4}
  \begin{align}
    &\tau_\rho + \beta\rho^2 X_\rho = 0\notag
    \\
    &\tau_\theta + 3\beta\rho^2 X_\theta = 0\notag
  \end{align}
\end{subequations}
which is equivalent to
\begin{subequations} \label{eq:4bis}
  \begin{align}
    &(\beta\rho^2 \tau + X)_\rho = (\beta\rho^2)_\rho\tau \notag
    \\
    &(3\beta\rho^2 \tau + X)_\theta = (3\beta\rho^2)_\theta\tau = 0\notag
  \end{align}
\end{subequations}
On solutions of~\eqref{52bis} the above system is fulfilled since it is
equivalent to the commutativity condition $[\phi,\psi]=0$ (see the Appendix). Algebraic solving system \eqref{eq:sol}
we obtain the implicit solutions:
\begin{equation}
  \label{eq:7}
  X = \frac{s_3}{2\beta\rho^3} - \frac{s_4'}{2\beta\rho},\quad
  \tau = -\frac{3s_3}{2\rho} - s_4 + \frac{\rho s_4'}{2}.
\end{equation}
Due to their implicit character, the solutions in \eqref{eq:7} will develop a
shock in finite time.

The Cauchy problem for \eqref{52bis} have been considered into details
in\cite{F}, but the possibility to find in a simple a direct way a large class
of exact smooth solutions seems to have been always skipped. Indeed, the main
theorem contained in \cite{F} speaks of \emph{piecewise smooth solutions
  with locally finitely many shocks}.  This kind of solutions are exactly the
ones represented by \eqref{eq:7}. On the other hand, we point out that Theorem
$2$\cite{F} speaks of \emph{soliton-like solutions}, i.e. a rotational wave
emerging as \emph{a travelling wave of unchanged shape}. This kind of solutions
seems to be exactly the ones we have explicitly determined in \eqref{eq:2}.

\section{Temple systems}

Let us put in the right framework the geometrical structure of system
\eqref{51}. To this end let us consider a general $2 \times 2$ autonomous and
uniform hyperbolic system
\begin{equation}\label{g1}
  u_t=[A(u,v)]_x, \quad u_t=[B(u,v)]_x,
\end{equation}
where $A$ and $B$ are smooth solutions. This system of differential equations
(for a given choice of $A$ and $B$) is defined in the base space
$(t,x)\times(u,v)$ prolonged to the jet space over this base space containing
the first order derivatives.  Our goal is to understand if in the subvariety of
the solutions of \eqref{g1}, say $\mathcal{S}$, it is possible a subset
$L\mathcal{S} \subset \mathcal{S}$, which may be determined solving a linear
differential equations.

This may be done using several methods, but in solving \eqref{51} we have used
the following one.  We have considered a subset of $(t,x)\times(u,v)$ defined
by a relation of the kind
\begin{equation}\label{g2}
  \varphi(u,v)=k,
\end{equation}
where $k$ is a constant. Then we have shown that when we restrict \eqref{g1} to
\eqref{g2} in the base space we obtain an overdetermined but compatible system
of equations which is linear.  Indeed since in the base space~\eqref{g2} holds
then whe should have in the jet space
$$
\varphi_u u_x+\varphi_v v_x=0, \quad \varphi_u u_t+\varphi_v v_t=0,
$$
and therefore from~\eqref{g1} (we impose $\varphi_v \neq 0$) we obtain the
overdetermined system
\begin{equation}\label{g3}
  u_t=\left( A_u-A_v \frac{\varphi_u}{\varphi_v}\right) u_x,
  \quad
  \frac{\varphi_u}{\varphi_v} u_t =
  \left(B_u-B_v \frac{\varphi_u}{\varphi_v}\right) u_x.
\end{equation}
By using the standard Lagrange-Charpit method we have that \eqref{g3} is fully
compatible if and only if
\begin{equation}\label{g4}
  B_u \varphi_v^2+(A_u-B_u) \varphi_u \varphi_v-A_v \varphi_u^2=0,
\end{equation}
and the single differential equation to which the overdetermined system is
reduced is linear, for example, if and only if
\begin{equation}\label{g5}
  A_u-A_v \frac{\varphi_u}{\varphi_v}=k,
\end{equation}
where we point out again that $k$ is constant.

The general solution of \eqref{g4} and \eqref{g5} is given by
\begin{subequations}
  \begin{align}
    & A(u,v)=H(\varphi)u+\Phi(\varphi),\\
    &B(u,v)=H(\varphi)v+\Psi(\varphi),
  \end{align}
\end{subequations}
where $H, \Phi$ and $\Psi$ are arbitrary functions of $\varphi(u,v)$ and we
have considered that we are solving the \eqref{g4} and \eqref{g5} in the
conical subset of the basic space defined by \eqref{g2}.

We have therefore determined a class of hyperbolic systems containing the
Temple system\cite{T,AF}
\begin{subequations}  \label{53}
  \begin{align} &u_t-[P(u,v)u]_x =0,
    \\
    & v_t-[P(u,v)v]_x=0,
  \end{align}
\end{subequations}
as special case.

For this system the corresponding Cauchy problem have been extensively
studied\cite{T} but once again we notice that the possibility to
deduce an infinity of solutions for \eqref{53} solving a linear equation seems
to have been skipped. Here we perform some general consideration about system
\eqref{53}. Let us rewrite \eqref{53} as
\begin{equation}\label{53bis}
  \left( \begin{array}{c}
      u_t \\
      v_t \end{array} \right) = \left( \begin{array}{cc}
      P_uu+P & P_v u \\
      P_u v & P_v v+P \end{array} \right) \left( \begin{array}{c}
      u_x \\
      v_x \end{array} \right) \end{equation}

The right-eigenvalues of $A$ in \eqref{53bis} are
\begin{equation}
  \lambda_1=P+P_uu+P_vv, \quad \lambda_2=P,
\end{equation}
and the corresponding right-eigenvectors are
\begin{equation}
  \vec{d}_1=(1, v/u), \quad \vec{d}_2=(1,-P_u/P_v).
\end{equation}

About \eqref{53bis} we remark:
\begin{itemize}
\item The eigenvector $\vec{d}_2$ is exceptional: indeed $\nabla \lambda_2 \cdot
  \vec{d}_2=0$.
\item We have $P(u,v)=P(u/v)$ if and only if $\lambda_1=\lambda_2$.
\item The system \eqref{53bis} is \emph{completely} exceptional if and only if
  $uP_u+vP_v=u^{-1}H(u/v)$, with $H$ arbitrary function.
\item The system \eqref{53bis} may be derived by a classical Hamiltonian
  density $\mathcal{H}[u]=\int H dx$ and rewritten as $u_t=[H_v]_x, \,
  v_t=[H_u]_x$ if and only if $vP_v=uP_u$, i.e. $P=P(uv)$.
\end{itemize}

Here we are interested in the case where only one eigenvalue is exceptional. In
this case we use only the Riemann invariant corresponding to this eigenvalue to
transform the system in a form where it is clear how to detect the class of
\emph{linear} equations. We introduce the point transformation of the dependent
variables $\alpha=\alpha(u,v)$ and $\beta=\beta(u,v)$ such that
$P(u,v)=R(\alpha)$ and $\beta=u/v$. In so doing \eqref{53} is rewritten in the
form
\begin{subequations}  \label{h3}
  \begin{align}
    &\alpha_t-(R'(\alpha_uu+\alpha_vv)+R(\alpha))\alpha_x =0,
    \\ & \beta_t - R(\alpha)\beta_x=0.
  \end{align}
\end{subequations}

This is a diagonal two-component hydrodynamic type system and, as such, it can
always be solved with the generalized hodograph method of the previous section.
However, for concrete cases of $R$ and $\alpha$ the system could be still
difficult to solve.

When $\alpha$ is constant (i.e. $P(u, v)$ is constant) the system \eqref{h3}
collapses to a single \emph{linear} differential equation.  Therefore it always
possible to find an infinity of \emph{linear} solutions for the nonlinear
system \eqref{53}.

In general the first equation~\eqref{h3} is of the form $\alpha_t -
f(\alpha,\beta)\alpha_x = 0$. However, it may happen that for special
functional forms of $R$ and $\alpha$ this equation decouples from the second
one. The condition for the decoupling is
\begin{multline}
  \label{eq:8}
  \pd{(\alpha_uu+\alpha_vv)}{\beta} =
0 \quad \Rightarrow \quad
\\
  \pd{u}{\beta}(\alpha_{uu}u + \alpha_{uv}v + \alpha_u) +
  \pd{v}{\beta}(\alpha_{uv}u + \alpha_{vv}v + \alpha_v) = 0.
\end{multline}
For instance, the above condition holds if $\alpha_{u}u + \alpha_{v}v$ is
constant; a function $\alpha$ fulfilling this last property $\alpha =
e^{a(u-v)+c}$ with $a$, $b$, $c$ three constants. Another possibility is that
$\alpha(u,v) = uv$.

The decoupled system is the general category to which the asymptotic system
associated with the elastic system \eqref{1} belongs. In this case
\begin{subequations} \label{h4}
  \begin{align}
    &\alpha_t-f(\alpha)\alpha_x =0,\\
    & \beta_t-R(\alpha)\beta_x=0.
  \end{align}
\end{subequations}

The same method as the asymptotic system applies to the above system. Indeed,
the hydrodynamic symmetries are
\begin{equation}
  \label{eq:351}
  \phi = s_1\alpha_x\frac{\partial}{\partial\alpha} +
  s_2\beta_x\frac{\partial}{\partial\beta}
\end{equation}
where $s_1=s_1(\alpha)$ is arbitrary and $s_2=s_2(\alpha,\beta)$ is subject to
the following linear ODE:
\begin{displaymath}
  R_\alpha(s_2 - s_1) + s_{2\,\alpha}(f - R) = 0
\end{displaymath}
which can be solved in a standard way. The resulting symmetries commute, hence
the generalized hodograph method provides the generic solution. For example, in
the particular case $\alpha(u,v)=uv$ we have $f(\alpha) = R_\alpha2\alpha + R$
and the above equation becomes $s_2 - s_1 + 2\alpha s_{2\,\alpha} = 0$ whose
explicit solution does not depend on $R$ and can be given in quadratures after
the form of $s_1$ is specified.

In the general situation when $f=f(\alpha,\beta)$ the constraint on $s_2$ in
the hydrodynamic symmetries is a partial differential equation whose solutions
must be investigated for each explicit form of $f$.

\begin{remark}
  The case when \eqref{53} is completely exceptional has been extensively
  studied\cite{BT}. In this case using Riemann invariants and then an hodograph
  transformation the system may be linearized.  A general theory for the
  linearization of completely exceptional second order hyperbolic conservative
  equations is provided in \cite{Donato}.  For example, let us consider
  the completely exceptional system obtained by setting $P(u,v)=u/v$. Now
  \eqref{h3} is
  \begin{equation} \label{h4a}
    u_t=\left(\frac{u^2}{v} \right)_x, \quad v_t=u_x,
  \end{equation}
  equivalent to
  \begin{equation} \label{h4bis} \psi_x=u, \quad \psi_t=\frac{u^2}{v},\quad
    \phi_x=v, \quad \phi_t=u.
  \end{equation}
  Being $\psi_x=\phi_t$ we have a \emph{stream} function $\chi$ such that
  $\psi=\chi_t, \quad \phi=\chi_x$ and the equation $\psi_t=u^2/v$ is therefore
  equivalent to the classical homogeneous Monge-Ampere equation $\chi_{tt}
  \chi_{xx}-\chi_{xt}^2=0,$ which, it is well known, is a linearizable
  equation.
\end{remark}

\section{Concluding Remarks}

We have considered the celebrated polarized circularly waves found by Carroll
in \cite{C}: this is a class of beautiful and simple general smooth exact
solutions for the nonlinear theory of isotropic elasticity. We have provided
evidence that the mathematical reason for the existence of such solutions is
the exceptional character of the hyperbolic system of determining equations for
such waves.  Indeed, we have been able to generalize such solutions in a
straightforward way.  To our knowledge, the huge class of exact solutions
(clearly obtainable by similarity methods) that we have obtained was not
noticed before, despite the fact that several papers have been devoted to the
group analysis of systems of wave equations in $1+1$ dimensions (see for
example \cite{V}).

To simplify the algebra we have restricted our attention to the asymptotic
first order system. For the system \eqref{1bis} the computation of the
generalized symmetries is a possible but cumbersome procedure. For this reason,
to determine similarity reductions the approach proposed by Carroll\cite{C1,C2}
used by Destrade and Saccomandi\cite{DS,DS2,DS3,DS1} via the
complex-coordinates formalism in \eqref{2} may be relevant.

For example, going back to the system of second order differential equations
\eqref{1bis} it is easy to understand how the Carroll's method may be extended.
Clearly this extension it is not trivial, elegant and beautiful as for system
\eqref{1}, but it is still an effective method to find by reduction to ordinary
differential equation some exact solutions of \eqref{1bis}.

A non trivial example of how it is possible to extend the ideas of Carroll is
given considering $P=P(uv)$ in \eqref{1bis}. In this case we start considering
$$u=\phi(t)\psi(x), \quad v=\frac{\phi(t)}{\psi(x)}.$$
In so doing we get $P=P(\phi^2)$ and we obtain the class of solution
$$u=\phi(t)\exp(kx), \quad v=\phi(t)\exp(-kx),$$
where $\phi_{tt}=k^2P(\phi^2)\phi$.

On the other hand, our discussion opens some interesting mathematical
questions.  First of all if there is the possibility of the existence of smooth
solutions (in the whole space) and this also for large initial data for the
equations of non-linear elasticity. Then if the global existence theorems by
Temple\cite{T} may be for certain conditions on the initial data reformulated
in this more strong setting. A first step in this direction seems given in
\cite{CRS}.

\textbf{Acknowledgments.} We would like to thank M.V. Pavlov for
helpful comments and suggestions. RV would also like to thank A.C. Norman for
his support with REDUCE. This work has been partially supported by Italian GNFM
of INdAM.

\section*{Appendix: symmetries and conservation laws}
\label{sec:append-symm-cons}

The system \eqref{52bis} is a hydrodynamic-type quasilinear system in diagonal
form and the coefficients of $\theta_\tau$ and $\rho_\tau$ are its Riemann
invariants \cite{}. However, the system is not completely exceptional \cite{}
(or linearly degenerate, according with another terminology \cite{}): the
gradients of its eigenvalues are not orthogonal to its eigenvectors.  An
entropy pair (or a conservation law) for \eqref{52bis} is given by any relation
\begin{displaymath}
  D_\tau T(\theta,\rho) - D_xX(\theta,\rho)=0,
\end{displaymath}
satisfied by all solutions of \eqref{h3}. (Here $D_\tau$ and $D_x$ are the
usual total derivatives).  The system admits the following hydrodynamic
conservation law densities:
\begin{equation}
  \label{eq:1}
  X= - \frac{dc_1}{d\rho} \rho^2 - 3 c_1 \rho - c_2 \rho ,\quad
  T=\beta\left( - 3 \frac{dc_1}{d\rho} r^4 - 3 c_1 \rho^3 - c_2 \rho^3\right),
\end{equation}
where $c_1$ is an arbitrary function of $\rho$ and $c_2$ is an arbitrary
function of $\theta$. The corresponding characteristic vector is the pair of
functions
\begin{align*}
  &\frac{\delta X}{\delta \theta} = - \frac{dc_2}{dv} \rho,
  \\
  &\frac{\delta X}{\delta\rho} = \left( - \frac{d^2c_1}{d\rho^2} \rho^2 - 5
    \frac{dc_1}{d\rho} \rho - 3 c_1 - c_2\right)
\end{align*}
where $\delta/\delta \theta$, $\delta/\delta\rho$ are variational derivatives.
The characteristic vector vanishes if and only if the above hydrodynamic
conservation laws are trivial, \emph{i.e.} they are the total divergence of a
quantity defined on the whole coordinate space. This happens if and only if
$c_2$ is a constant and $c_1$ fulfills the above Cauchy--Euler ODE. So, the
space of nontrivial hydrodynamic conserved quantities is still parametrized by
two functions which are `almost' arbitrary.

The system \eqref{52bis} also admits the following symmetries whose
characteristic function depends on first-order derivatives:
\begin{equation}
  \label{eq:3}
  \phi = - s_1 \frac{\partial}{\partial\theta} +
  s_2 \rho_\tau\frac{\partial}{\partial\rho}
\end{equation}
where $s_1=s_1(\theta,\rho,\theta_\tau)$ and $s_2=s_2(r)$ fulfill the
additional PDE
\begin{displaymath}
  \frac{ds_1}{d\rho} \rho + \frac{ds_1}{d\theta_\tau} \theta_\tau + s_2
  \theta_\tau = 0.
\end{displaymath}
If we require that $\phi$ be of hydrodynamic type \cite{GSW}, \emph{i.e.}
$s_1=s_{10}(\rho,\theta)\theta_\tau$ then we conclude that
$s_{10}=s_3(\theta)(1/\rho) + s_4(\rho)$ and $s_2 = - (ds_4/d\rho\cdot \rho +
s_4)$. The hydrodynamic symmetries $\phi$ commute with the vector field whose
characteristic function is given by the right-hand side of the equation. More
precisely, if
\begin{displaymath}
  \psi = \beta \rho^2\theta_\tau \frac{\partial}{\partial \theta} +
  3\beta \rho^2 \rho_\tau \frac{\partial}{\partial \rho},\quad
  \phi = -\left(\frac{s_3}{\rho} + s_4\right)\theta_\tau
  \frac{\partial}{\partial \theta}
  -\left(\frac{ds_4}{d\rho}\rho + s_4\right)\rho_\tau
  \frac{\partial}{\partial \rho}
\end{displaymath}
then we have
\begin{align*}
  \{\phi,\psi\} = & \left( \phi^\theta\pd{\psi^\theta}{\theta} +
    \phi^\rho\pd{\psi^\theta}{\rho} +
    D_\tau(\phi^\theta)\pd{\psi^\theta}{\theta_\tau} + D_\tau
    \phi^\rho\pd{\psi^\theta}{\rho_\tau}\right.
  \\
  & \left. - \psi^\theta\pd{\phi^\theta}{\theta} -
    \psi^\rho\pd{\phi^\theta}{\rho} -
    D_\tau(\psi^\theta)\pd{\phi^\theta}{\theta_\tau} - D_\tau
    \psi^\rho\pd{\phi^\theta}{\rho_\tau} \right)\pd{}{\theta} +
  \\
  & \left( \phi^\theta\pd{\psi^\rho}{\theta} + \phi^\rho\pd{\psi^\rho}{\rho} +
    D_\tau(\phi^\theta)\pd{\psi^\rho}{\theta_\tau} + D_\tau
    \phi^\rho\pd{\psi^\rho}{\rho_\tau}\right.
  \\
  & \left. - \psi^\theta\pd{\phi^\rho}{\theta} - \psi^\rho\pd{\phi^\rho}{\rho}
    - D_\tau(\psi^\theta)\pd{\phi^\rho}{\theta_\tau} - D_\tau
    \psi^\rho\pd{\phi^\rho}{\rho_\tau} \right)\pd{}{\rho} = 0,
\end{align*}
where $\{\phi,\psi\}$ is the Jacobi bracket (see, for example, \cite{Many}), or
the commutator of the two flows $\psi$ and $\phi$.

We stress that the same computation can be repeated for the system \ref{h4}.



\begin{thebibliography}{99}

\bibitem{AF} S.I. Agafonov, E.V. Ferapontov: Systems of conservation laws
    of Temple class, equations of associativity and linear congruences in
    $\mathbf{P}^4$. Manuscripta Math. \textbf{106} (2001), no. 4, 461--488.

\bibitem{Ant} S.S. Antman, Nonlinear Problems of Elasticity.  Springer Verlag,
  New York, 1995.

\bibitem{Ball} J. Ball, Some open problems in elasticity, 3--59 in Geometry,
  Mechanics, and Dynamics edited by P. Newton, P. Holmes and A.  Weinstein
  (2002) Springer New York

\bibitem{Many} A. V. Bocharov, V. N. Chetverikov, S.V.  Duzhin, N.G. Khorkova,
  I.S.  Krasilshchik, A.V.  Samokhin, Yu.N. Torkhov, A.M. Verbovetsky and A. M.
  Vinogradov: Symmetries and Conservation Laws for Differential Equations of
  Mathematical Physics, I.S. Krasilshchik and A. M.  Vinogradov eds.,
  Translations of Math. Monographs 182, Amer. Math. Soc.  (1999).

\bibitem{Boi1} G. Boillat, T. Ruggeri, Su alcune classi di potenziali
  termodinamici come conseguenza dell'esistenza di particolari onde di
  discontinuit\'a nella meccanica dei continui con deformazioni
  finite. Rend. Sem.  Matematico Univ. di Padova, vol. LI (1974) 293--304.

\bibitem{BT} G. Boillat, T. Ruggeri, Characteristic shocks: Completely and
  strictly exceptional systems, Boll. Un.  Mat. Ital. 15-A (1978) 197--204.

\bibitem{Boi2} G. Boillat, S. Pluchino, Onde eccezionali in mezzi iperelastici
  con deformazioni finite piane. J. Appl. Math. Phys. (ZAMP) 35 (1984)
  363--372.

\bibitem{CRS} C. Carassa, M. Rascle, D. Serre, \'Etude d'un mod\'ele
  hyperbolique en dynamique des cables RAIRO M2AN 19 (1985) 573--599.

\bibitem{C} M.M. Carroll, Some Results on Finite Amplitude Elastic Waves.  Acta
  Mechanica 3 (1967) 167--181.

\bibitem{C1} M. M. Carroll, Oscillatory shearing of nonlinearly elastic solids.
  Z. angew. Math. Phys. (ZAMP) 25 (1974) 83--88.

\bibitem{C2} M. M. Carroll, Plane elastic standing waves of finite
  amplitude. J. Elast. 7 (1977) 411--424.

\bibitem{Collins} W. D. Collins, One Dimensional Nonlinear Wave Propagation in
  Incompressible Elastic Materials, Q. J. Mech. and Appl. Maths. 19, 259--328
  (1966).

\bibitem{Dafermos} C. M. Dafermos, Hyperbolic Conservation Laws in Continuum
  Mechanics, Springer-Verlag 1995, 2005.

\bibitem{DS} M. Destrade, G. Saccomandi, Some results on finite amplitude
  elastic waves propagating in rotating media, Acta Mechanica, 173 (2004)
  19--31.

\bibitem{DS2} M. Destrade, G. Saccomandi, On finite amplitude elastic waves
  propagating in compressible solids, Physical Review E, 72 (2005) 016620.

\bibitem{DS3} M. Destrade, G. Saccomandi, Solitary and compact-like shear waves
  in the bulk of solids, Physical Review E, 73 (2006) 065604.

\bibitem{DS1} M. Destrade, G. Saccomandi, Nonlinear transverse waves in
  deformed dispersive solids, Wave Motion 45 (2008) 325--336.

\bibitem{Donato1} A. Donato, Legge di evoluzione delle discontinuit\'a e
  determinazione di una classe di potenziali elastici compatibile con la
  propagazione di onde eccezionali in un mezzo continuo sottoposto a
  particolari deformazioni finite. J. Appl. Math. Phys. (ZAMP) 28 (1977)
  1059--1066.

\bibitem{Donato} A. Donato, F.  Oliveri, Linearization of completely
  exceptional second order hyperbolic conservative equations, Applicable
  Analysis: An International Journal 57: (1995) 35--45.

\bibitem{GSW} A.M. Grundland, M.B. Sheftel, P. Winternitz, Invariant solutions
  of equations of the hydrodynamic type, J.Phys.A, 33(46):8193--8215, 2000.

\bibitem{E1} D.G. Ebin, Global solutions of the equations of elastodynamics of
  incompressible neo- Hookean materials. Proc. Nat. Acad. Sci. U.S.A.,
  90:3802--3805, 1993.

\bibitem{E2} D.G. Ebin, Global solutions of the equations of elastodynamics for
  incompressible materials, Electron. Res. Announc. Amer. Math. Soc., 2:50Ð59
  (electronic), 1996.

\bibitem{E3} D.G. Ebin, R.A. Saxton, The initial value problem for
  elastodynamics of incompressible bodies. Arch. Rational Mech. Anal.,
  94:15--38, 1986.

\bibitem{F} H. Freist\"uhler, On the Cauchy problem for a class of hyperbolic
  systems of conservation laws, J. of Diff. Equations, 112:170--178. 1994.

\bibitem{Renard} W.J. Hrusa, M. Renardy, An existence theorem for the Dirichlet
  problem in the elastodynamics of incompressible materials, Arch. Rational
  Mech. Anal., 102:95--117, 1988. Corrections ibid 110:373-375,1990.

\bibitem{KK} B. Keyfitz and H. Kranzer, A system of non-strictly hyperbolic
  conservation laws arising in elasticity theory, Arch. Rat. Mech. Anal. 72
  (1980) 219--241.

\bibitem{Kla} S. Klainerman and A. Majda, Formation of singularities for wave
  equations including the nonlinear vibrating string, Communications on Pure
  and Applied Mathematics XXXIII (1980) 241--263.

\bibitem{Ros3} V. Krylov, P. Rosenau Solitary waves in an elastic string
  Original Research Article Physics Letters A 217 (1996)31--42.

\bibitem{JT} A. Jeffrey, M. Teymur, Formation of shock waves in hyperelastic
  solids, Acta Mechanica 20 (1974) 133--149.

\bibitem{Fritz} F. John, Formation of singularities in one-dimensional
  nonlinear wave propagation, Communications on pure and applied mathematics
  XXVII (1974) 377--405.

\bibitem{Lei0} Hin-Chi Lei, Ming-Jui Hung Linearity of waves in some systems of
  non-linear elastodynamics International Journal of Nonlinear Mechanics 32
  (1997) 353--360.

\bibitem{Lei} Zhen Lei, T. C. Sideris, Yi Zhou,
Almost Global Existence for 2-D Incompressible Isotropic Elastodynamics.
to appear Transaction A.M.S. 2014.

\bibitem{Liu} T. P.Liu and C.H. Wang On a nonstrictly hyperbolic system of
  Conservation Laws, J.of differential equations 5 (1985) 1--14.

\bibitem{Olver} P.J. Olver, Applications of Lie Groups to Differential
  Equations, Springer 1992.

\bibitem{reduce} REDUCE, a computer algebra system; freely available at
  Sourceforge: \url{http://reduce-algebra.sourceforge.net/}

\bibitem{Rogers} C. Rogers, On a Coupled Nonlinear Schr\"odinger System: A
  Ermakov Connection, Studies in Applied Mathematics, to appear 2013.

\bibitem{Ros1} P.  Rosenau, M.B. Rubin, Motion of a nonlinear string: Some
  exact solutions to an old problem, Phys. Rev. A 31 (1985) 3480--3482.

\bibitem{Ros2} P.  Rosenau, M.B. Rubin, Some nonlinear three-dimensional
  motions of an elastic string, Physica D: Nonlinear Phenomena 19 (1986)
  433--439.

\bibitem{Scott} N.H. Scott, Acceleration waves in incompressible elastic solids
  Q. J. Mech.\ Appl.\ Math.\ XXIX (1976) 295--310.

\bibitem{Sideris} T. C. Sideris, B. Thomases, Global existence for
  three-dimensional incompressible isotropic elastodynamics, Communications on
  Pure and Applied Mathematics, LX (2007) 1707--1730.

\bibitem{T} B. Temple, Global Solution of the Cauchy Problem for a Class of
  $2x2$ Nonstrictly Hyperbolic Conservation Laws, Advances in Appl. Math. 3
  (1982) 335--375.

\bibitem{ST} S.P. Tsarev, The geometry of Hamiltonian systems of hydrodynamic
  type. The generalized hodograph method. Math.\ USSR Izvestya Vol.\ 37 no.\ 2
  (1991), 397--419.

\bibitem{V} P. J. Vassiliou, Coupled systems of nonlinear wave equations and
  finite-dimensional lie algebras I Acta Appl. Math. 8 (1987) 107--147

\bibitem{cdiff} R. Vitolo, CDIFF: a Reduce package for computations in
  the goemetry of differential equations, software, user guide and examples
  freely available at \texttt{http://gdeq.org}.

\end{thebibliography}
\end{document}